\documentclass[reprint,superscriptaddress,amsmath,amssymb,aps,pre]{revtex4-1}

\usepackage{graphicx}
\usepackage{dcolumn}
\usepackage{bm}
\usepackage{color}
\usepackage{bbold}
\usepackage{soul}

\usepackage[margin=8mm,textfont=small,
position=top,
labelformat=empty,
caption=false,
justification=raggedright, singlelinecheck=false]{subfig}

\providecommand*{\threeDD}{\textsc{3D}}

\usepackage{prettyref}
\newrefformat{sec}{Sec.~\ref{#1}}
\newrefformat{app}{Appendix~\ref{#1}}
\newrefformat{fig}{Fig.~\ref{#1}}
\newrefformat{tab}{Tab.~\ref{#1}}
\newrefformat{eq}{Eq.~(\ref{#1})}

\usepackage{mathtools} 
\usepackage{braket}

\newcommand{\overh}{\ensuremath{1/h}}

\newcommand{\Ione}{\ensuremath{I_1}}
\newcommand{\Itwo}{\ensuremath{I_2}}

\newcommand{\Irsone}{\ensuremath{I_{\text{res,1}}}}
\newcommand{\Irstwo}{\ensuremath{I_{\text{res,2}}}}

\newcommand{\Ms}{\ensuremath{M_{\text{res}}}}
\newcommand{\Vr}{\ensuremath{V_{\bm r}}}
\newcommand{\Vs}{\ensuremath{V_{\bm s}}}

\newcommand{\thetaone}{\ensuremath{\theta_1}}
\newcommand{\thetatwo}{\ensuremath{\theta_2}}

\newcommand{\vecI}{\ensuremath{\bm I}}

\newcommand{\vecIs}{\ensuremath{\bm I_{\text{res}}}}
\newcommand{\vectheta}{\ensuremath{\bm \theta}}

\newcommand{\vecresr}{\ensuremath{\bm r}}
\newcommand{\vecress}{\ensuremath{\bm s}}

\newcommand{\vecIm}{\ensuremath{\vecI_{\bm m}}}

\newcommand{\hamilton}{\ensuremath{H ( \vectheta, \vecI ) }}

\newcommand{\potential}{\ensuremath{V ( \vectheta ) }}

\newcommand{\hnode}{\ensuremath{H_0  }}
\newcommand{\hnodeI}{\ensuremath{H_0 ( \vecI ) }}

\newcommand{\Pabs}{\ensuremath{\hat{P}_{\Lambda}}}

\newcommand{\vecIop}{\ensuremath{\hat{\vecI}}}
\newcommand{\vecthetaop}{\ensuremath{\hat{\vectheta}}}

\newcommand{\potentialop}{\ensuremath{V ( \hat{\vectheta} ) }}

\newcommand{\unpertubedop}{\ensuremath{H_0 ( \hat{\vecI} ) }}

\newcommand{\psiketi}{\ensuremath{\ket{\psi_{\bm 0}}}}

\newcommand{\psabs}{\ensuremath{\Lambda}}

\newcommand{\Iabs}[1]{\ensuremath{\Lambda_{#1}}}

\newcommand{\Hquant}{\ensuremath{\hat{H}}}

\newcommand{\Ham}{Hamiltonian}

\DeclarePairedDelimiter\abs{\lvert}{\rvert}
\DeclarePairedDelimiter\norm{\lVert}{\rVert}%

\newcommand{\Ibaseket}[1]{\ensuremath{\ket{\vecI_{#1}}}}
\newcommand{\Ibasebra}[1]{\ensuremath{\bra{\vecI_{#1}}}}

\newcommand{\tunpath}{\ensuremath{\Gamma}}
\newcommand{\weight}[1]{\ensuremath{w_{\bm #1}}}

\newcommand{\fourfour}{\ensuremath{4 \times 4}}
\newcommand{\Hblock}{\ensuremath{\Hquant_{4 \times 4}}}

\newcommand{\oneD}{\textsc{1d}}
\newcommand{\twoD}{\textsc{2d}}
\newcommand{\threeD}{\textsc{3d}}
\newcommand{\fourD}{\textsc{4d}}

\newcommand{\sixD}{\textsc{6d}}

\newcommand{\onetori}{\oneD-tori}

\newcommand{\twotori}{\twoD-tori}

\newcommand{\poincare}{Poincar\'e}

\newcommand{\ps}{phase space}

\usepackage{color}

\usepackage[nice]{nicefrac}

\usepackage{prettyref}
\newrefformat{sec}{Sec.~\ref{#1}}
\newrefformat{app}{Appendix~\ref{#1}}
\newrefformat{fig}{Fig.~\ref{#1}}
\newrefformat{tab}{Tab.~\ref{#1}}
\newrefformat{eq}{Eq.~(\ref{#1})}

\usepackage{catchfile}
\newcommand{\getenv}[2][]{%
\CatchFileEdef{\temp}{"|kpsewhich --var-value #2"}{\endlinechar=-1}%
\if\relax\detokenize{#1}\relax\temp\else\let#1\temp\fi}

\getenv[\articlestoragepath]{ARTICLE_STORAGE}
\IfFileExists{pdf_article_link.sty}{\usepackage{pdf_article_link}}

\makeatletter
\let\Hy@backout\@gobble
\makeatother

\newcommand{\HIDDEN}[1]{}

\begin{document}

\title{Resonance--assisted tunneling in 4D normal--form Hamiltonians}

\author{Markus Firmbach}
\affiliation{Technische Universit\"at Dresden, Institut f\"ur Theoretische
             Physik and Center for Dynamics, 01062 Dresden, Germany}
\affiliation{Max-Planck-Institut f\"ur Physik komplexer Systeme, N\"othnitzer
    Stra\ss{}e 38, 01187 Dresden, Germany}

\author{Felix Fritzsch}
\affiliation{Technische Universit\"at Dresden, Institut f\"ur Theoretische
             Physik and Center for Dynamics, 01062 Dresden, Germany}

\author{Roland Ketzmerick}
\affiliation{Technische Universit\"at Dresden, Institut f\"ur Theoretische
             Physik and Center for Dynamics, 01062 Dresden, Germany}
\affiliation{Max-Planck-Institut f\"ur Physik komplexer Systeme, N\"othnitzer
Stra\ss{}e 38, 01187 Dresden, Germany}

\author{Arnd B\"acker}
\affiliation{Technische Universit\"at Dresden, Institut f\"ur Theoretische
             Physik and Center for Dynamics, 01062 Dresden, Germany}
\affiliation{Max-Planck-Institut f\"ur Physik komplexer Systeme, N\"othnitzer
Stra\ss{}e 38, 01187 Dresden, Germany}

\date{\today}

\begin{abstract}
    Nonlinear resonances in the
    classical phase space lead to a significant enhancement
    of tunneling.
    We demonstrate that the double resonance
    gives rise to a complicated tunneling peak structure.
    Such double resonances occur in Hamiltonian systems
    with an at least four-dimensional phase space.
    To explain the tunneling peak structure, we use the universal description
    of single and double resonances by
    \fourD{} normal--form Hamiltonians.
    By applying perturbative methods, we reveal the underlying mechanism
    of enhancement and suppression of tunneling
    and obtain excellent quantitative agreement.
    Using a minimal matrix, we obtain model an intuitive understanding.
\end{abstract}

\pacs{PACS here}

\maketitle

\section{Introduction}\label{sec:intro}

Quantum tunneling connects classically disjoint regions
and therefore is one of the most prominent features of quantum mechanics.
In particular, the description of tunneling through energy barriers
allowed for computing of molecular ground states \cite{Hun1927}
and explaining radioactive decay \cite{Gam1928, GurCon1929}.
However, classical barriers may arise not only from potential
barriers but can be generated by the classical dynamics
in phase space leading to the concept of
dynamical tunneling \cite{DavHel1981, KesSch2011}.
The paradigmatic example for this is tunneling between dynamically
unconnected
regions of a mixed phase space in which chaotic and regular motion coexists.
There, for instance, tunneling between two regular regions,
separated by chaotic dynamics, is moderated by chaos--assisted tunneling
through the chaotic
component of phase space \cite{LinBal1990, BohTomUll1993, TomUll1994}.
The  tunneling between regular and chaotic regions is
described by regular--to--chaotic tunneling
\cite{LoeBaeKetSch2010, SchMouUll2011}.
Being a fundamental quantum mechanical effect, dynamical tunneling
is of relevance in many different fields of physics,
e.g.\ vibrational spectra
of molecules \cite{DavHel1981, Hel1999, Kes2007},
systems of ultra cold atoms \cite{Hen2001, SteOskRai2001},
optical microcavities
\cite{ShiHarFukHenSasNar2010, ShiHarFukHenSunNar2011,
YanLeeMooLeeKimDaoLeeAn2010, YiKulKimWie2017, YiKulWie2018,
KulWie2016b, KwaShiMooLeeYanAn2015},
and microwave billiards
\cite{DemGraHeiHofRehRic2000, BaeKetLoeRobVidHoeKuhSto2008,
    DieGuhGutMisRic2014},
and explains power-law level repulsion at small energy
spacings in systems with mixed phase spaces \cite{BaeKetLoeMer2011}.

The presence of nonlinear resonances
can  drastically enhance the tunneling
between disconnected regions in phase space
by the mechanism of resonance--assisted tunneling
\cite{Ozo1984, BroSchUll2001, BroSchUll2002}.
Under variation of a parameter
significant enhancement of tunneling is observed. This was
recently demonstrated in experiments
for microwave resonators \cite{GehLoeShiBaeKetKuhSto2015}
and optical microcavities \cite{KwaShiMooLeeYanAn2015}
by varying the frequency or the shape of the boundary.
A lot of progress has been made concerning the theoretical understanding
of resonance--assisted tunneling
\cite{
    Ozo1984, BroSchUll2001, BroSchUll2002,
    ShuIke1995, ShuIke1998, PodNar2003, Kes2003, PodNar2005, EltSch2005,
    Kes2005b, SheFisGuaReb2006, Kes2007, BaeKetLoeSch2008,
    BaeKetLoeRobVidHoeKuhSto2008, ShuIke2008, ShuIshIke2008, ShuIshIke2009a,
    ShuIshIke2009b, BaeKetLoeWieHen2009, BaeKetLoe2010,
    LoeBaeKetSch2010, MerLoeBaeKetShu2013, HanShuIke2015, ShuIke2016, KulWie2016b,
    MerKulLoeBaeKet2016, FriBaeKetMer2017}
mostly concentrating on systems
effectively described by two-dimensional (\twoD{}) maps.
One of the key tools for the theoretical description
is the mapping of the system with nonlinear resonances to a
universal
pendulum-like \twoD{} normal--form Hamiltonian,
arising from secular perturbation theory
\cite{Ozo1984, BroSchUll2002} or normal form theory \cite{LebMou1999}.
The quantized normal--form Hamiltonians allow us to explain
the universal features of resonance--assisted tunneling
in \twoD{} quantum maps respectively,
making them a key ingredient for a comprehensive understanding.

Dynamical tunneling and specifically resonance--assisted tunneling has also
been studied in higher--dimensional systems
\cite{Cre1994, Kes2005, Ric2012, PitTanHel2016, KarKes2018}.
Generically in such systems resonances of higher rank arise,
which are not present in \twoD{} systems.
The case of rank 2 is called double resonance and is the simplest
case showing new types of dynamics.
It occurs in an at least four-dimensional
symplectic map or \sixD{} Hamiltonian.
In the vicinity of a double resonance the dynamics is
effectively described
by the interpolating flow of
a \fourD{} normal--form Hamiltonian
\cite{WalFor1969, Tod1994, Hal1997, Hal1999, GelSimVie2013}.
Thus, studying tunneling in such \fourD{} normal--form Hamiltonians
provides a first step towards the understanding of
resonance--assisted tunneling in \fourD{} symplectic maps.
Experimentally realizable systems with double resonances are, for example
three-dimensional optical microcavities or microwave resonators
as well as periodically driven two-dimensional systems.
Moreover, the quantized normal--form Hamiltonians
have applications in the studies
of vibrational dynamics of chemical molecules
\cite{SwiDel1979, UzeMil1991, Hel1999, Gru2004,
    SemKes2003, SemKes2004, KesEzr1997,
    Kes2007, FarSchGuoJoy2009, Kes2013c, Lei2015}.
In particular, the intramolecular energy transfer is heavily influenced
by classical nonlinear resonances, as they allow for couplings
between different vibrational modes \cite{Hel1999, Kes2007}.
However, a quantitative description based on resonance--assisted tunneling
has not been worked out so far.

In this paper we give a qualitative as well as a quantitative
description of resonance-assisted tunneling
in the presence of a double resonance.
To this end, we consider the classical dynamics of the simplest
\fourD{} normal--form Hamiltonians,
which describe the dynamics of either a single
coupled resonance or a double resonance.
Subsequently we study the phase-space localization of the eigenstates
of the quantized systems in classically disjoint regions.
We find that tunneling in the vicinity of the single resonance shows the
same characteristics as in \twoD{} systems, i.e., a drastic enhancement of
tunneling.
Such an enhancement is also seen in the case of the double resonance,
but occurs in a very complex fashion.
Additionally, we also observe the suppression of tunneling.
Utilizing a perturbative expansion
based on paths in action-space
we are able to quantitatively explain the complicated peak structure
of enhancement as well as suppression.
This provides a first step toward a
detailed understanding of resonance--assisted
tunneling in higher dimensions.
The application to generic \fourD{} quantum maps is beyond
the scope of this paper and remains subject of further research.

The paper is organized as follows:
In \prettyref{sec:rat}, a brief sketch of
the general approach is given.
By means of the instructive example of tunneling in the
case of a single coupled resonance, we
introduce the methods and notation in Sec.~\prettyref{subsec:scr}.
Section~\prettyref{subsec:doubleres} treats
the central case of a double resonance.
For both cases, the classical normal-form model
as well as the associated quantum system is investigated.
Quantitatively resonance-assisted tunneling is studied by means of numerical
diagonalization and a perturbative description of the underlying
mechanism is presented.
The key features of resonance-assisted tunneling
in the normal--form Hamiltonians can be intuitively understood within a
minimal \fourfour{} matrix model presented in
Sec.~\prettyref{subsec:matrixmodel}.
Finally, a summary and outlook is given in \prettyref{sec:summary}.

\section{Resonance assisted tunneling}\label{sec:rat}

Classical nonlinear resonances are ubiquitous in Hamiltonian dynamical systems
and manifest themselves by means of resonance-assisted tunneling in the
corresponding quantum system. In order to explain the quantum features
based on classical properties,
we utilize a universal description of the classical
dynamics in terms of a truncated \fourD{} normal--form Hamiltonian
approximating both
single and double resonances
\cite{WalFor1969, Tod1994, Hal1997, Hal1999, GelSimVie2013}.
The corresponding \fourD{} \ps{} is separated into
dynamically disjoint regions by the resonance channels associated with the
nonlinear resonances. The eigenstates of the quantum
Hamiltonian predominantly localize on classical quantizing tori,
located in one of these regions.
However, due to tunneling, there
is also non--vanishing probability in the other, classically forbidden,
regions. We introduce the weight in these regions as a quantitative measure of
resonance-assisted tunneling which shows
similar characteristics as known from two dimensional quantum maps
\cite{MerKulLoeBaeKet2016}.
A perturbative description of resonance-assisted tunneling is obtained
for both single coupled resonances and double resonances.
This allows for an intuitive understanding of the enhancement and,
in the case of
double resonances, also the suppression of tunneling in terms of perturbative
paths along a discrete action grid.

\subsection{Single coupled resonance}\label{subsec:scr}
First the instructive case of the single coupled resonance
in a \fourD{} Hamiltonian system is considered,
which will turn out to be quite similar to \twoD{} systems
with a single dominating resonance.
To this end, we start with the classical resonant Hamiltonian
\cite{WalFor1969, Tod1994, Hal1997, Hal1999, GelSimVie2013}
\begin{equation}\label{eq:scr_cl_ham}
\hamilton = \hnodeI + \potential
\end{equation}
obtained from normal form analysis and truncation. It is expressed
in action-angle coordinates
\( {\vecI = \left( I_1, I_2 \right) }\)
and
\( { \vectheta = \left(\theta_1, \theta_2\right) }\)
of \hnodeI.
Keeping only the non-resonant, quadratic order, gives
\begin{equation}\label{eq:scr_cl_ham_h0}
\hnodeI = \frac{1}{2}
\left(\vecI - \vecIs\right)^{\text{T}}\mathbf{M}
\left(\vecI - \vecIs \right)
\end{equation}
for some real symmetric matrix $\mathbf{M}$,
such that $\hnodeI$ has an isolated extreme at $\vecIs$.

For the single resonance the lowest order resonant term is given by
$ \potential = 2\Vs \cos \left( \bm s \vectheta \right) $
and the resonance vector $ \vecress = \left( s_1, s_2 \right) $.
By means of a canonical transformation either the resonance vector can be
brought to the form
$ \vecress = \left( s_1, 0 \right) $, i.e.\ an uncoupled resonance, or
$\mathbf{M}$ may be transformed to diagonal form. Here, we choose
$\mathbf{M} = \Ms^{-1}\mathbf{1}$.
Since the potential \potential{} is $2\pi$-periodic the \ps{}
of the system is the cylinder
$\mathbb{T}^{2} \times \mathbb{R}^2 $.
Furthermore, the system is integrable because energy is conserved and
$ s_2\Ione - s_1 \Itwo $ is a second constant of motion called
polyad number \cite{Kel1990}.

\begin{figure*}
    \includegraphics{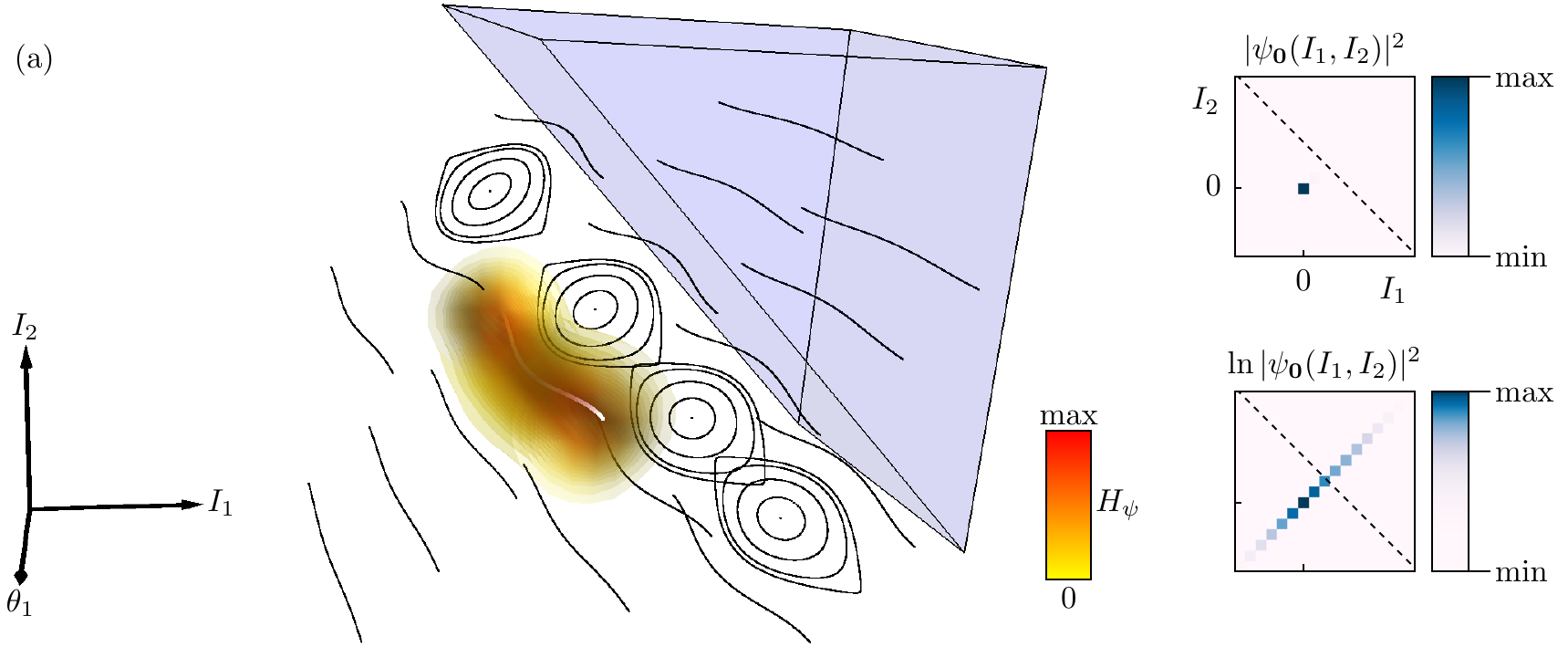}\\
    \includegraphics{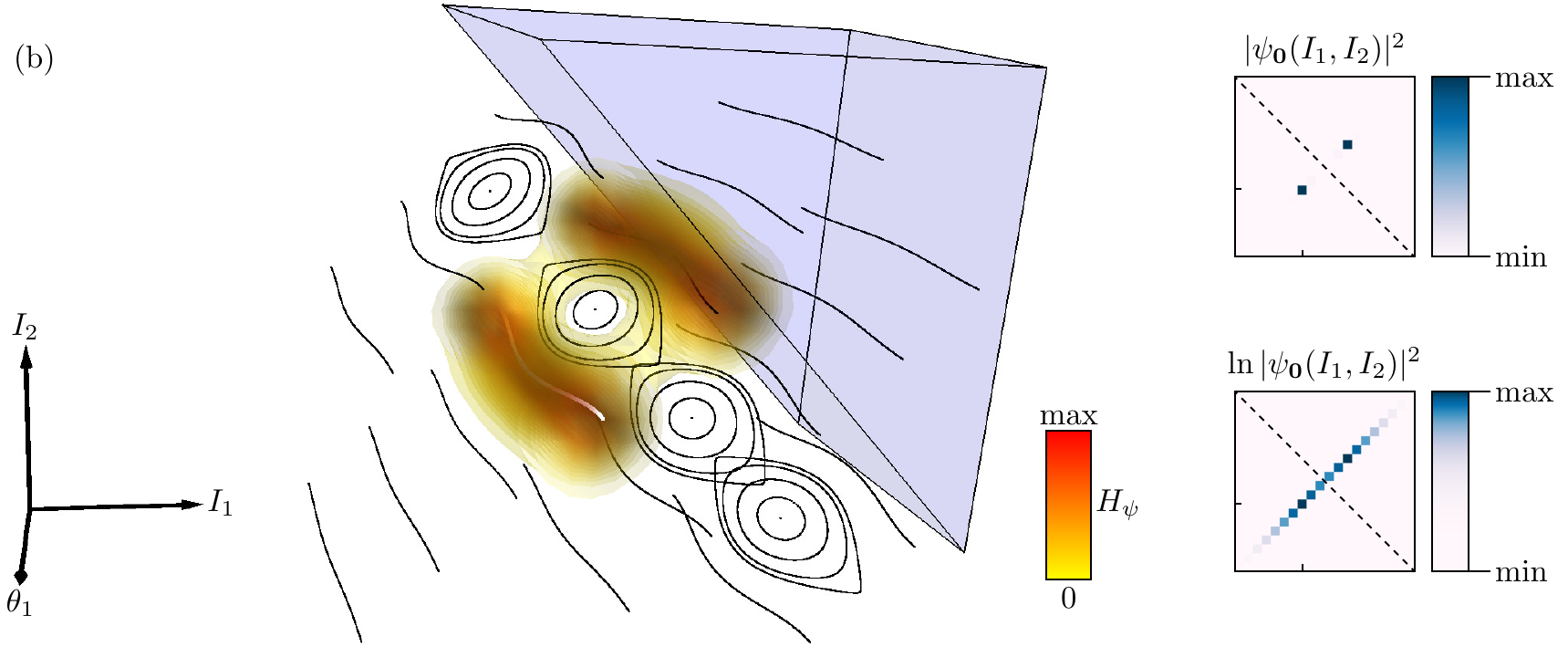}
    \caption{
        Phase space of the single coupled resonance shown in a
        \threeD{} hyperplane for $\theta_2 = 0$ fixed.
        Regular structures are shown as \oneD{} black curves
        for four planes with constant polyad number.
        The eigenstate $\ket{\psi_{\bm 0}}$ maximally
        localizing on the classical
        quantizing torus $\vecI_{\bm 0} = \left( 0, 0 \right)$
        (white line) is shown in a \threeD{} Husimi
        phase-space representation (normal scale, see color bar)
        in (a) for $\overh = 0.3450$ and
        (b) for $\overh = 0.3979$.
        The region $\psabs$, \prettyref{eq:scr_abs},
        trivially extended in the $\theta_1$-direction,
        is shown in light blue.
        For comparison the squared amplitude of the eigenstate
        $|\psi_{\bm 0}(I_1, I_2)|^2$
        is shown  in action-space representation
        (normal and logarithmic scale).
        The resonance center line, \prettyref{eq:resonance_line},
        is shown as dashed line.
        }
    \label{fig:scr}
\end{figure*}

For concrete numerical calculations and visualization we choose
$\bm s = (1,1)$, $\Vs=0.1$,
$\Ms = 0.8$ and $\vecIs=(1.0, 1.0)$.
Instead of
restricting the \fourD{} phase space to a \threeD{} energy manifold
and introducing a \poincare{} section to obtain a \twoD{}
representation of the dynamics,
we show the \threeD{} hyperplane for a fixed value of
$\thetatwo = 0$ and
visualize the remaining three coordinates $(\thetaone, \Ione, \Itwo)$.
This allows for an intuitive representation
of the phase-space structures for different
energies.
In \prettyref{fig:scr} this hyperplane for $\thetatwo = 0$ is shown.
Regular dynamics takes place on \twotori{} so that they
appear as \oneD{} curves. Furthermore,
the \fourD{} \ps{} is foliated into invariant \threeD{} planes
of constant polyad number.
In the \threeD{} hyperplane they appear as \twoD{} planes
resembling the dynamics of a pendulum.
In \prettyref{fig:scr}, orbits in four such planes are shown.
The stable and unstable fixed points of these pendulums correspond to elliptic
and hyperbolic \onetori{} in the full phase space. Their projection onto the
action coordinates is called resonance center line
\cite{LicLie1992} and fulfills
\begin{equation}\label{eq:resonance_line}
s_1 \frac{\partial \hnodeI}{\partial \Ione}
+
s_2  \frac{\partial \hnodeI}{\partial \Itwo}
= 0.
\end{equation}
The families of the elliptic and hyperbolic \onetori{}
build up the skeleton of the resonance channel \cite{OnkLanKetBae2016}.
The regular structures form a tilted tube in phase space as
shown in \prettyref{fig:scr}.
Note, that the geometry of the \fourD{} normal form system,
represented in the section with $\theta_2=0$,
is very similar to that of a corresponding
\fourD{} symplectic map, represented in a
\threeD{} phase space slice \cite{RicLanBaeKet2014}.

The quantum system is obtained from
the classical \Ham{} \prettyref{eq:scr_cl_ham}
by means of Weyl quantization \cite{Gro1946}, which yields the operators
\( \vecthetaop  \) and
\( \vecIop  \)
as well as the Hamiltonian
 $ \hat{H} =\unpertubedop{} + \potentialop{}$.
The eigenvalue equation
\begin{equation}\label{eq:eigeneq}
\hat{H} \ket{\psi_{\bm m}} = E \ket{\psi_{\bm m}}
\end{equation}
gives the eigenstates and eigenenergies of the system.

The periodicity of \potentialop{}
implies, analogously to Bloch's theorem \cite{Blo1929},
the discretization of action space,
in terms of the grid
\begin{equation}\label{eq:action_space}
\vecIm = \hbar \left(\bm m + \bm \vartheta \right)
\quad \text{with} \quad
\bm m \in \mathbb{Z}^2.
\end{equation}
Here, $\hbar$ denotes an effective Planck's constant, which may be defined as
the ratio of Planck's constant and a typical action of the system and plays
the role of a semiclassical parameter, i.e.\ $\hbar \to 0$ corresponds to the
semiclassical limit.
Note that changing the inverse effective Planck constant is experimentally
possible, e.g.,
in a microwave cavity by varying the frequency
\cite{GehLoeShiBaeKetKuhSto2015}.
Furthermore, we fix the Bloch phase $ \bm\vartheta = (0 ,0) $.
Evaluation of the Hamiltonian in the action basis \prettyref{eq:action_space}
gives the matrix elements
\begin{equation}
\hat{H}_{\bm m, \bm n}
=
\hnode \left( \vecI_{\bm n} \right) \delta_{\bm m, \bm n}
+
\Vs \left( \delta_{\bm m, \bm n + \bm s} +
           \delta_{\bm m, \bm n - \bm s} \right).
\end{equation}
As we are interested in quantum states in the vicinity of the resonance
channel near $\vecI = (0, 0)$,
we restrict the action grid to a rectangle, outside which the wave
functions are supposed to carry negligible probability.
Truncation at $| \Ione |, | \Itwo | \leq 10 $ gives rise to a
finite-dimensional Hilbert space.

Note, that \unpertubedop{} is diagonal in action space
and therefore its eigenstates $\Ibaseket{\bm m}$ are labeled
by the quantum number
$\bm m \in \mathbb{Z}^2$.
By means of semiclassical Einstein-Brillouin-Keller (EBK)
quantization, this relates them to
the quantizing torus $\vecIm$, \prettyref{eq:action_space},
of the classical system given by $\hnodeI$.
As long as \potentialop{} is a small perturbation of
\unpertubedop{} this association of quantum numbers remains
valid and the eigenstates of interest localize on  quantizing
tori of the perturbed \Ham{} with the same action as
$\vecI_{\bm m}$.
Our choice of
$ \bm \vartheta = (0 ,0) $
as well as $\vecIs$, $\Ms$, and $\Vs$ ensures the existence of an eigenstate
$\ket{\psi_{\bm 0}} \equiv \ket{\psi_{\bm m}}$
with quantum number $\bm m = (0, 0)$ which localizes on the
corresponding quantizing torus
$\vecI_{\bm 0} = \left( 0, 0 \right)$ close to, but still outside of, the
resonance channel.
In \prettyref{fig:scr} this torus is depicted as a white line.
The choice $ \bm m = (0, 0)$ is convenient as it allows us to study quantum
states associated with the same torus when varying $\hbar$.

In order to visualize the eigenstates $\ket{\psi}$ in phase space, we
use the Husimi representation \cite{Hus1940}
\begin{equation} \label{eq:husimi_func}
H_{\psi} \left(\vectheta, \vecI \right) = \frac{1}{\hbar^{2}}
\abs{\Braket{ \alpha_{\text{coh}} \left( \vectheta,  \vecI \right) |
        \psi }} ^2,
\end{equation}
defined by the overlap of $\ket{\psi}$ with a coherent state
$\ket{\alpha_{\text{coh}}  \left( \vectheta,  \vecI \right)}$ of minimal
uncertainty centered at $ \left( \vectheta,  \vecI \right)$
given in action representation as
\begin{equation} \label{eq:coh_state}
\Braket{\vecI |
    \alpha_{\text{coh}} \left( \vectheta_{0},  \vecI_{0}\right)} =
\frac{1}{\sqrt{\pi \hbar}}
\exp \left( - \frac{ \left(\vecI - \vecI_{0} \right)^2}{2\hbar}
+ \frac{\text{i}}{\hbar} \vecI \vectheta_{0}
\right),
\end{equation}
and considered on the torus \cite{NonVor1998}.

By fixing $\theta_2=0$, the Husimi representation can be
compared with the classical phase space structures.
Note that this representation is similar to the \threeD{} Husimi
representation on the \threeD{} phase space slice \cite{RicLanBaeKet2014}.
In \prettyref{fig:scr}(a) the eigenstate $\ket{\psi_{\bm 0}}$
for $\overh = 0.3450$ is shown. The Husimi representation is
localizing on the quantizing torus $\vecI_{\bm 0}$.
This can also be seen in the action-space representation
as one point with high intensity in the normal scale
and exponential tails along a diagonal in action space,
as can be seen in the logarithmic scale.

The effect of resonance-assisted tunneling can be measured
in various ways.
Following the studies of regular--to--chaotic tunneling
or partially open quantum systems
we measure the weight of $\ket{\psi_{\bm 0}}$ on the opposite side of the
resonance $\bm s = (s_1, s_2)$, accounting for the geometry
of phase space \cite{MerKulLoeBaeKet2016}.
In order to determine this weight we formally follow
Ref.~\cite{MerKulLoeBaeKet2016} and introduce a
region $\psabs$ in action space
given by
\begin{align}\label{eq:scr_abs}
\begin{split}
 \psabs \coloneqq
\lbrace &
\left( \Ione, \Itwo \right)
\in \mathbb{R}^2 : \\
&\Itwo \geq - \frac{s_1}{s_2}\left(\Ione - \Irsone \right) + \Irstwo + \Iabs{1}
\rbrace .
\end{split}
\end{align}
The trivial extension of \psabs{} in the $\theta_1$--direction
is shown as blue shaded region in \prettyref{fig:scr}.
The boundary is chosen parallel to the resonance center line.
It is located on the opposite side of the resonance channel with respect to
the quantizing torus
$\vecI_{\bm 0} = \left( 0, 0 \right)$.
Its distance from the
resonance channel is controlled by $\Iabs{1}$.
As $\psabs$ is dynamically separated from the quantizing torus
$\vecI_{\bm 0}$,
the weight of the state $\ket{\psi_{\bm 0}}$ on $\psabs$ measures the strength
of tunneling over the resonance channel.
For the quantum system this translates into the projector
\cite{MerKulLoeBaeKet2016}
\begin{equation} \label{eq:def_absorber}
\Pabs \ket{\vecI} = \chi_{\Lambda} \left( \vecI \right) \ket{\vecI},
\end{equation}
where $\chi_{\Lambda}$ denotes the characteristic function,
\begin{equation}
\chi_{\Lambda}\left( \vecI \right) = \begin{cases}
1 & \text{for} \; \vecI \in \psabs  \\
0 & \text{for} \; \vecI \notin \psabs .
\end{cases}
\end{equation}
The weight $\weight{0}$ of $\ket{\psi_{\bm 0}}$
in \psabs{} is determined by
\begin{equation}\label{eq:tunneling_rate}
\weight{0} = \norm{\Pabs \psiketi}^2.
\end{equation}
Note, that opening the system in $\Lambda$ leads
to decay rates showing qualitatively the same results
as the weights $\weight{0}$ \cite{MerKulLoeBaeKet2016}.
In experiments, the weight can be related to quality factors
of optical modes in
microcavities or to decay rates of resonance states.

\begin{figure}
    \includegraphics{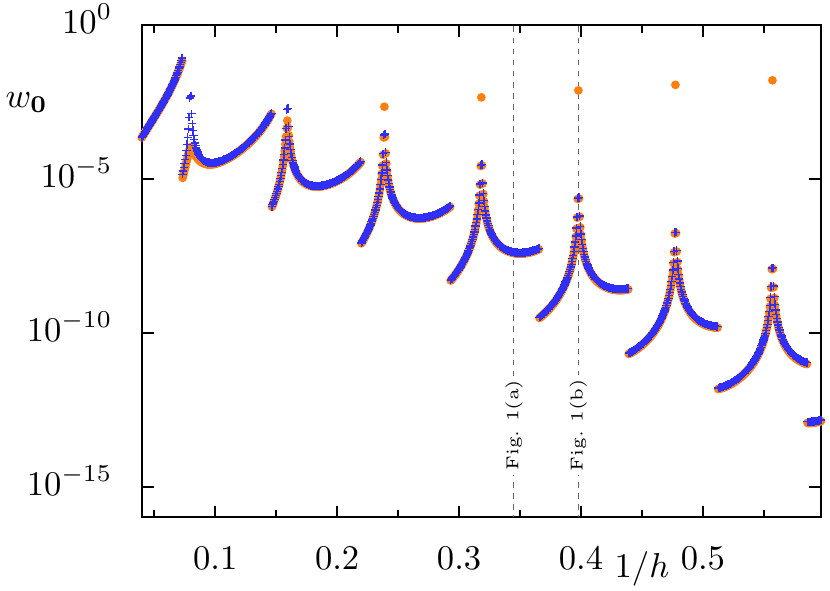}
    \caption{Weight $\weight{0}$ of state $\ket{\psi_{\bm 0}}$,
        see \prettyref{eq:tunneling_rate},
        as function of the inverse effective Planck constant \overh{}
        for the single coupled resonance shown as orange bullets.
        The perturbative prediction \eqref{eq:scr_short}
        is shown as blue crosses.
    }
    \label{fig:scr_rates}
\end{figure}

In \prettyref{fig:scr_rates} the weight $\weight{0}$
is depicted semilogarithmically as a function of the inverse effective
Planck constant \overh{} as orange bullets.
On average it shows an exponential decay
with increasing \overh{},
i.e.\ when approaching the semiclassical limit $\hbar \to 0$.
This overall trend is well known from \twoD{} quantum maps
\cite{BroSchUll2002, BaeKetLoe2010}
and experiments \cite{GehLoeShiBaeKetKuhSto2015}.
In addition to the overall exponential decay there are quantization
jumps at equidistant values of \overh{}.
They are due to points of the discrete action grid that pass the border of
the region \psabs{} when $h$ is varied.
That is, the quantization jumps are caused by the choice of
$\psabs{}$ and the associated projector.
In principle, the jumps could be avoided by smoothing the border of
the region \psabs{}.
However, as these quantization jumps are not related to tunneling
we keep the definition of the projector
\prettyref{eq:def_absorber} for simplicity.
In the numerical computation, we choose $\Iabs{1}=2.35$, which ensures
that quantization jumps and peaks do not coincide
in the considered range of \overh{}.

Most importantly,
on top of the overall exponential decay there are also sharp peaks
occurring in a regular manner
at which tunneling is drastically enhanced by several orders of magnitude.
These peaks are the manifestation of resonance--assisted tunneling first
discussed in \twoD{} systems \cite{BroSchUll2001, BroSchUll2002}.
There, the enhancement of tunneling was traced back to degenerate
eigenstates of the unperturbed Hamiltonian,
localizing on quantizing tori symmetrically
located with respect to the resonance channel.
Therefore, also for the \fourD{} system,
right on top of one of the peaks, the state
$\ket{\psi_{\bm 0}}$
is expected to localize equally on the torus
$\vecI_{\bm 0}$
and its symmetric counterpart
in the plane of the same polyad number.
This is nicely seen in the
Husimi distribution of $\ket{\psi_{\bm 0}}$ for $\overh = 0.3979$ and
$\thetatwo = 0$ fixed
and the corresponding action-space representation
shown in \prettyref{fig:scr}(b).
In contrast away from the peak the symmetric partner does not
belong to the discrete action grid.

In order to gain insight into the underlying mechanism of
resonance-assisted tunneling in \fourD{} normal--form systems we follow
Refs.~\cite{BroSchUll2001, BroSchUll2002} and use a
perturbative approach to give a quantitative accurate description
of the weights.
To this end we start with \unpertubedop{} and the orthonormal basis
of action states
$\ket{\vecIm}$,
in which \unpertubedop{} is diagonal, and consider \potentialop{} as
a sufficiently small perturbation.
The only non-vanishing matrix elements of \potentialop{} occur between
action states whose quantum numbers differ by $\pm \bm s$
\begin{equation}\label{eq:scr_coupling}
\Ibasebra{\bm m} \potentialop \Ibaseket{\bm m + \bm s} =
\Vs.
\end{equation}
The perturbative expansion of a state localizing predominantly
on the quantizing torus
$\vecI_{\bm m} $
is thus given by \cite{LoeBaeKetSch2010, SchMouUll2011}
\begin{equation}\label{eq:scr_pt_state}
\ket{\psi^{\text{pert}}_{\bm m}} = \Ibaseket{\bm m}
+ \sum_{l \in \mathbb Z \setminus \{0\}}
A_{\bm m}^{(l)} \Ibaseket{\bm m + l\bm s}.
\end{equation}
As the polyad number has a quantum mechanical
analogue given by the polyad number operator
$\hat{n} = s_2\hat{\Ione} - s_1 \hat{\Itwo}$ commuting with $\hat{H}$
only states within the subspace of same polyad number contribute to
this perturbative expansion.
For fixed $l$ the coefficient $A_{\bm m}^{(l)}$ can be decomposed into
distinct contributions
$\lambda_{\bm m}^{\tunpath}$,
that differ in the order of perturbation theory.
Each contribution is uniquely associated with a sequence
$\tunpath = \left[\bm t_1, \bm t_2, \ldots , \bm t_k \right]$ of couplings
with $\bm t_i \in \{\pm \bm s\}$ which
defines a path on the action grid.
The length of the path is given by the number of elements in the
sequence and is denoted as $|\tunpath | := k$. It
coincides with the order of perturbation theory in
which $\tunpath$ contributes.

A path can contribute to $A_{\bm m}^{(l)}$
only if $|\tunpath | \geq |l|$.
Graphically such a  path $\tunpath$ connects
$\vecI_{\bm m}$ with
$\vecI_{\bm m + l \bm s}$ on the discrete action grid,
where intermediate subsequent actions
differ by $\pm \bm s$ in their quantum numbers.

\begin{figure}
    \includegraphics{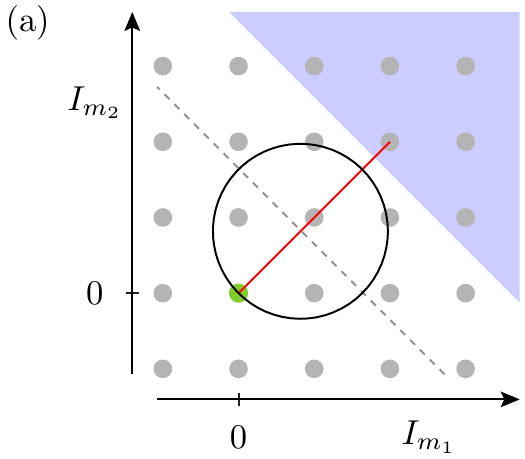}
    \includegraphics{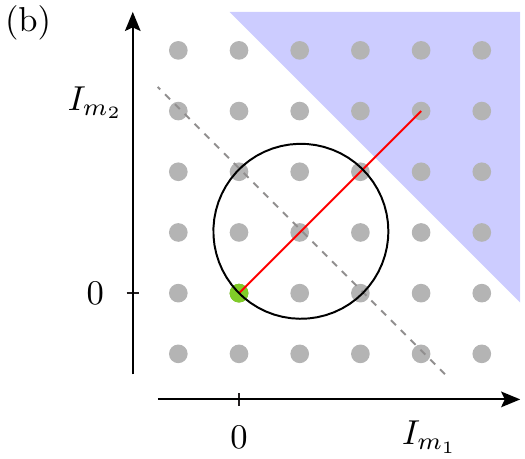}
    \caption{Scheme of perturbation theory for the single coupled
        resonance.
        The action grid \( \vecIm \) is depicted
        as gray points, the
        quantizing torus $\vecI_{\bm 0} = \left( 0, 0 \right)$
        as green point.
        The dashed line shows the position of the resonance center line,
        see \prettyref{eq:resonance_line}.
        The circle indicates the level set of constant energy
        $\hnode(\vecI_{\bm 0})$.
        The blue shaded area shows the region \psabs{},
        see \prettyref{eq:scr_abs}.
        The red line indicates the shortest path
        $\tunpath$.
        In (a) in the non-resonant case for $ \overh = 0.1314 $
        and $|\tunpath | = 2 $.
        In (b) the resonant case for \( \overh = 0.1638 \)
        and $|\tunpath | = 3 $.
        }
    \label{fig:scr_pert}
\end{figure}

Schematically this is illustrated in \prettyref{fig:scr_pert},
where the action grid is indicated by gray points
and the region \psabs{} is shaded blue.
Figure~\ref{fig:scr_pert}(a)
shows the path $\tunpath = [\bm s, \bm s]$
of length $|\tunpath| = 2$ connecting $\vecI_{\bm 0}$,
marked as green point, with $\vecI_{(2, 2)}$.
In \prettyref{fig:scr_pert}(b)
the path  $\tunpath = [\bm s, \bm s, \bm s]$ of length
$|\tunpath| = 3$ ending on $\vecI_{(3, 3)}$
is shown.

Let $\mathcal{M}_{\bm m}^{l}$ denote the set of all paths which connect
$\vecI_{\bm m}$ with $\vecI_{\bm m + l \bm s}$ and for which no
intermediate action coincides with $\vecI_{\bm m}$, i.e.\ paths coming
back to their starting point are excluded in the following
according to perturbation theory \cite{Loe1951}.
Each path $\tunpath \in \mathcal{M}_{\bm m}^{l}$
then contributes with
\begin{equation}\label{eq:scr_pt_prod_coeff}
\lambda_{\bm m}^{\tunpath}  = \Vs^{\abs{\tunpath}}
\prod_{i=1}^{\abs{\tunpath}}
\frac{1}{ \hnode \left(\vecIm \right) -
    \hnode \left( \vecI_{\bm m + \sum_{j=1}^{i} \bm t_j }  \right) },
\end{equation}
which includes the unperturbed energies of the intermediate actions.
Finally, the contributions from different paths add up to the coefficients
\begin{equation}\label{eq:scr_coeffs}
A_{\bm m}^{(l)} =
\sum_{\tunpath \in \mathcal{M}_{\bm m}^{l}}\lambda_{\bm m}^{\tunpath}.
\end{equation}
Inserting the state $\ket{\psi^{\text{pert}}_{\bm m}} $,
\prettyref{eq:scr_pt_state} into \prettyref{eq:tunneling_rate} we find
for the weight in the region \psabs{}
\begin{equation}\label{eq:scr_pt_rate}
\weight{m} = \sum_{\substack{l \in \mathbb N \\
\vecI_{\bm m + l\bm s} \in \psabs}}
\big | A_{\bm m}^{(l)} \big |^2 .
\end{equation}

From the \twoD{} case, it is known that it suffices to take only the
unique shortest path
$\tunpath_l=\left[ \bm s , \bm s, \ldots \bm s \right]$
of length $l$ into account.
The sum in \prettyref{eq:scr_pt_rate} may thus be rewritten as
a sum over all positive integers $l$ starting
from the smallest integer $l_{\text{min}}$ for which
$\vecI_{\bm m + l_{\text{min}}\bm s} \in \psabs$
holds and where we approximate
$A_{\bm m}^{(l)} \approx \lambda_{\bm m}^{\tunpath_l}$.
In particular, neglecting all but the lowest contributing order
$l_{\text{min}}$
of perturbation theory, i.e.\ including only the shortest
path, \prettyref{eq:scr_pt_rate}
reduces to
\begin{equation}\label{eq:scr_short}
\weight{m} = \big | \lambda_{\bm m}^{\tunpath_{l_{\text{min}}}}\big |^2 .
\end{equation}
Note that the paths shown in \prettyref{fig:scr_pert} are exactly
the shortest for the two different values of $1/h$, respectively.
This lowest order approximation already leads
to excellent agreement with the weights $\weight{0}$
obtained from diagonalization
of the Hamiltonian, see \prettyref{fig:scr_rates},
where the prediction \prettyref{eq:scr_short} is shown as blue crosses
comparing very well with the numerical results. This is because
higher orders are suppressed by $\abs{\Vs}$, i.e.\ by at least one order
of magnitude.

The emergence of resonance--assisted tunneling peaks
can be traced back to the energy denominators
in \prettyref{eq:scr_pt_prod_coeff}.
The contribution $\lambda_{\bm m}^{\tunpath}$
diverges whenever an intermediate state
$\Ibaseket{\bm m + l \bm s}$ and $\Ibaseket{\bm m}$
are resonant, i.e.\ they are energetically degenerate
with respect to \unpertubedop{},
which gives rise to the peaks.
This mechanism is well known from \twoD{} systems.
It is further illustrated in \prettyref{fig:scr_pert}(a), where the
non-resonant case is shown.
There no intermediate state lies on the circular level set of constant energy
$\hnode(\vecI)=\hnode(\vecI_{\bm 0}) = \text{constant}$.
In contrast \prettyref{fig:scr_pert}(b) depicts the resonant case where
both $\vecI_{\bm 0}$ and $\vecI_{(2, 2)}$
have approximately the same unperturbed energy.
Note, that in the case of exact degeneracy of
$\vecI_{\bm 0}$ with an intermediate
state, the perturbative prediction diverges and thus looses its validity.
This situation, which in principle could be treated by degenerate
perturbation theory, is also ignored in the work on \twoD{} system.
Furthermore, quantization jumps occur in
the perturbative result as well and coincide with the values of \overh{}
at which the length $l_{\text{min}}$ of the shortest path into the
region \psabs{} changes by one.
Despite being a \fourD{} system due to integrability and the both classically
and quantum mechanically conserved polyad number  \( s_2 \Ione - s_1\Itwo \),
the results closely resemble what is known from \twoD{} systems.

\subsection{Double resonance}\label{subsec:doubleres}

Resonances of higher rank are possible in higher--dimensional systems only.
The minimal example is the double resonance in a \fourD{} system.
Using normal-form theory and truncation yields an effective
Hamiltonian
\cite{WalFor1969, Tod1994, Hal1997, Hal1999, GelSimVie2013}
\begin{equation}\label{eq:doubleres_ham}
\hamilton = \hnodeI + 2V_{\bm r} \cos(\bm r \vectheta)
+ 2V_{\bm s} \cos(\bm s \vectheta),
\end{equation}
which describes the dynamics in the vicinity of a double resonance,
occurring at the isolated minimum of $\hnodeI$.
Using \prettyref{eq:scr_cl_ham_h0} for $\hnodeI$ this minimum
occurs at $\vecI = \vecIs$.
Choosing linear independent resonance vectors $\bm r, \bm s$ gives rise to two
different resonance channels parametrized by the corresponding
resonance center line, see \prettyref{eq:resonance_line}.
At the intersection of both resonance center lines,
i.e.~the minimum of $\hnodeI$,
the double resonance condition is fulfilled.

In general, \prettyref{eq:doubleres_ham} gives rise to non-integrable dynamics,
except for specific choices of $\bm r, \bm s$
\cite{GelSimVie2013}.
In particular, due to the second resonant term the polyad
numbers corresponding to either of the two resonance vectors are no
longer conserved.
However, whenever perturbations are small the system still can be considered
as near-integrable.

\begin{figure*}
    \includegraphics{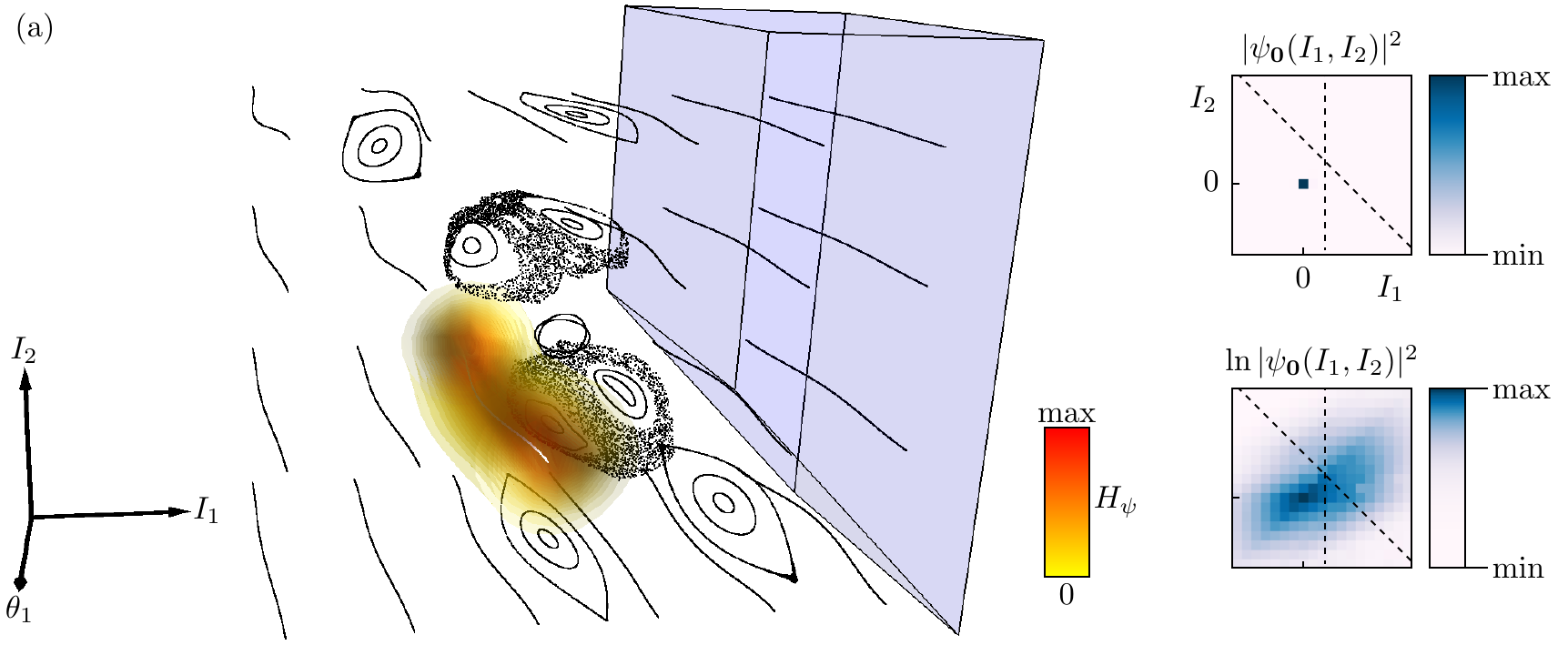}
    \includegraphics{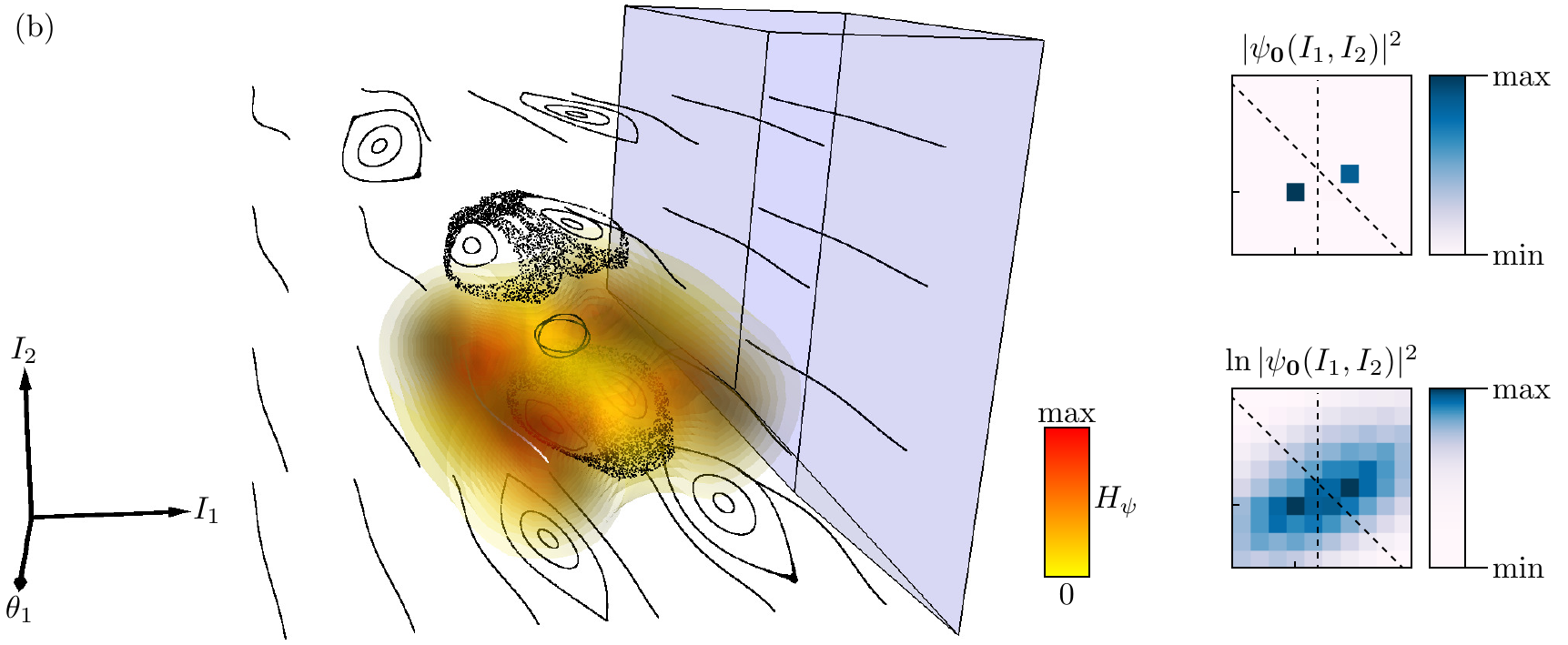}
    \includegraphics{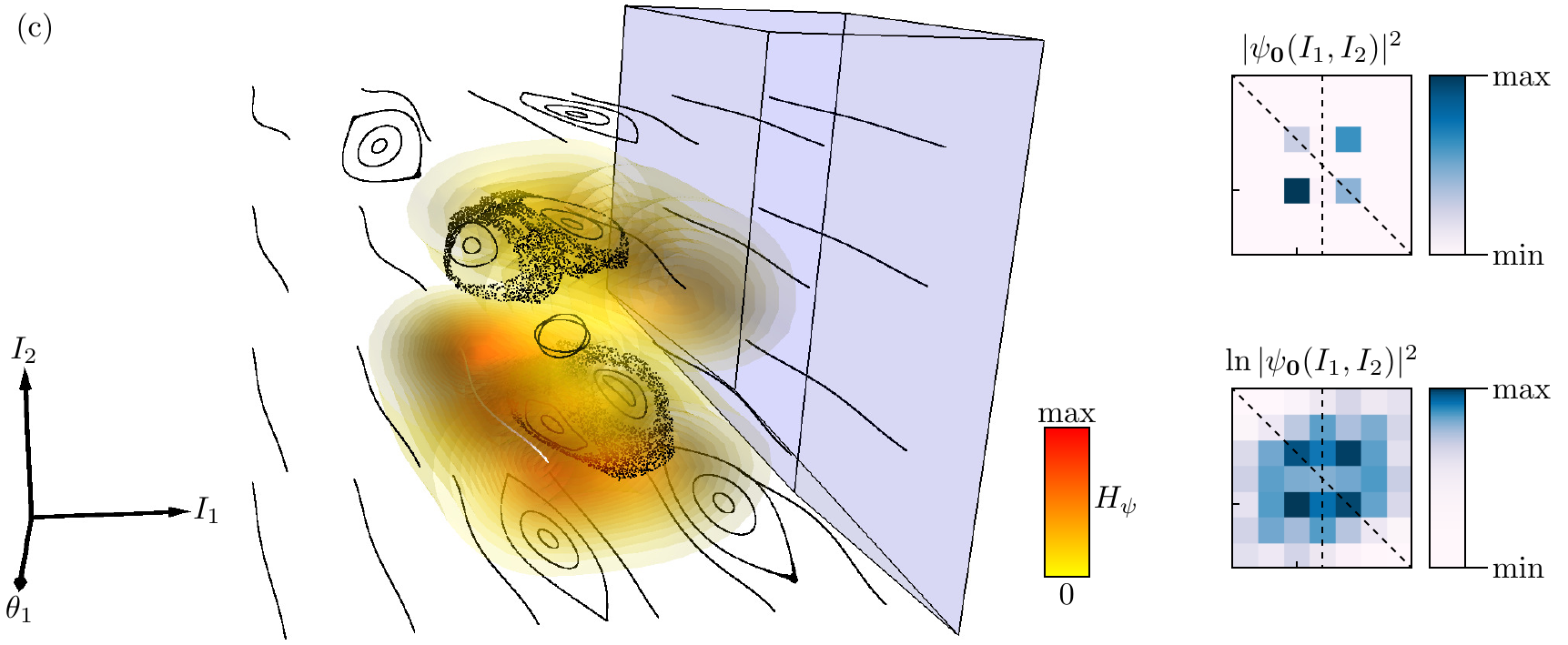}
    \caption{
        Phase space of the double resonance shown in a
        \threeD{} hyperplane for $\theta_2 = 0$ fixed.
        Regular structures are
        shown as \oneD{} black curves, whereas chaotic structures are shown
        as points.
        The eigenstate $\ket{\psi_{\bm 0}}$ maximally
        localizing on the classical
        quantizing torus $\vecI_{\bm 0} = \left( 0, 0 \right)$
        (white line) is shown in a \threeD{} Husimi
        phase-space representation (normal scale, see colorbar)
        in (a) for $\overh = 0.375$,
        (b) for $1/h = 0.1991$, and
        (c) for  $1/h = 0.1591$.
        The region $\psabs$, \prettyref{eq:double_res_abs},
        trivially extended in the $\theta_1$-direction,
        is shown in light blue.
        For comparison, the squared amplitude of the eigenstate
        $|\psi_{\bm 0}(I_1, I_2)|^2$
        is shown  in action-space representation
        (normal and logarithmic scale).
        The resonance center lines, \prettyref{eq:resonance_line},
        are shown as dashed lines.
}
    \label{fig:double_res}
\end{figure*}

In the following we choose
$ \vecresr = \left( r_1, r_2 \right) = \left( 1, 0 \right) $
and
$ \vecress = \left( s_1, s_2 \right) = \left( 1, 1 \right) $
as linearly independent resonance vectors with
prefactors $\Vr = \Vs=0.05$ and
$\hnodeI$ as given by \prettyref{eq:scr_cl_ham_h0}
for $\Ms = 1$ and $\vecIs=(1.0, 1.0)$.
The phase space is shown in the \threeD{} hyperplane
for fixed $\thetatwo=0$
in \prettyref{fig:double_res}.
Both resonance vectors give rise to a resonance channel
intersecting in a so-called resonance junction at the minimum
of \hnodeI{} at $\vecIs = \left( 1.0, 1.0 \right)$.
This corresponds to the point in which the double
resonance condition is fulfilled.
It gives rise to
four equilibria with different types of stability, namely
one elliptic-elliptic at \( (\theta_1, \theta_2) =  (\pi, 0) \),
one hyperbolic-hyperbolic at \( (0, 0) \), and
and two elliptic-hyperbolic at \( (0, \pi) \) and
\( (\pi, \pi) \) respectively \cite{HowMac1987, Tod1996}.
Together with their invariant manifolds or attached families of elliptic
\oneD{} tori they organize the phase space close to $\vecIs$.
Along the resonance center line of each resonance and
sufficiently far away from $\vecIs$, the
phase space locally resembles the phase space of a \twoD{} pendulum and
is governed by the associated resonance channel.
Following the resonance channels towards the junction at $\vecIs$ chaotic
layers of both resonance channels begin to overlap and form a
connected stochastic layer.
As usual in non-integrable systems also resonances of higher order occur.
These regular resonance structures are not shown, but are indirectly seen
by the small holes in the chaotic layers.

\begin{figure}
    \includegraphics{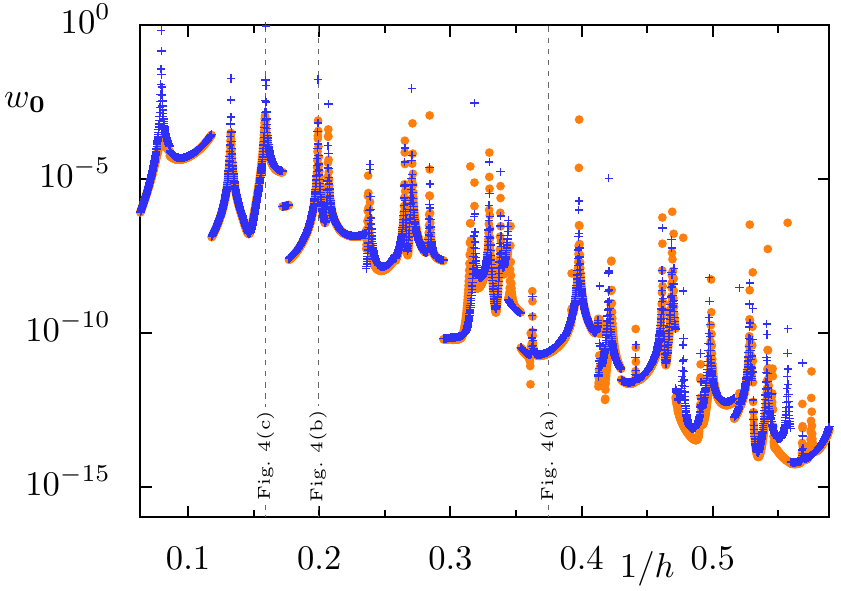}
    \caption{
        Weight $\weight{0}$ of state $\ket{\psi_{\bm 0}}$,
        see \prettyref{eq:tunneling_rate},
        as function of the inverse effective Planck constant \overh{}
        for the double resonance is shown as orange bullets.
        The perturbative prediction \eqref{eq:doubleres_pt_rate},
        only using the shortest paths,
        is shown as blue crosses.
}
    \label{fig:rates_double_res}
\end{figure}

From \prettyref{fig:double_res}, we conclude that phase space and
action space are each divided into four dynamically separated regions by
the resonance channels and the resonance center lines, respectively.
Therefore, tunneling across both of the resonance channels is expected.
As the chaotic layer does not support a relevant number of chaotic
eigenstates in the considered regime of $\hbar$, we do not expect it
to influence tunneling.

Accounting for the additional second resonant term
in \prettyref{eq:doubleres_ham},
the matrix elements of the quantized
Hamiltonian read
\begin{align}
\hat{H}_{\bm m, \bm n}
=
\hnode \left( \vecI_{\bm n} \right) \delta_{\bm m, \bm n}
&+
\Vs \left( \delta_{\bm m, \bm n + \bm s} +
\delta_{\bm m, \bm n - \bm s} \right) \nonumber\\
&+
\Vr \left( \delta_{\bm m, \bm n + \bm r} +
\delta_{\bm m, \bm n - \bm r} \right).
\end{align}
Our choice of parameters guarantees the existence of a
quantizing torus $\vecI_{\bm 0}$ for all values of
the effective Planck's constant.
This allows for studying tunneling of the eigenstate $\ket{\psi_{\bm 0}}$
in terms of its weight in a properly chosen region
\psabs{} in action space
located on opposite sides of both resonance channels.
We choose
\begin{align}\label{eq:double_res_abs}
\begin{split}
\psabs \coloneqq \lbrace &
\left( \Ione, \Itwo \right)
\in \mathbb{R}^2 : \\
& \Itwo \geq - \frac{s_1}{s_2}\left(\Ione - \Irsone \right) + \Irstwo
+ \Iabs{1}, \\
& \Ione \geq \Iabs{2} \rbrace .
\end{split}
\end{align}
For numerical computations we set $\Iabs{1} = 1.7$ and $\Iabs{2}=2.7$.
In \prettyref{fig:double_res} the region \psabs{}
is shown as a light blue region, trivially extended in the
$\theta_1$--direction.
Using \psabs{} in the definition of the projector \prettyref{eq:def_absorber},
the weight $\weight{0}$ is again given by
\prettyref{eq:tunneling_rate}.

The weight $\weight{0}$ of $\ket{\psi_{\bm 0}}$ on $\psabs$
is shown in
\prettyref{fig:rates_double_res}
as a function of the inverse effective Planck constant $1/h$ by orange bullets.
Besides the overall exponential decay of the weight in the
region \psabs{}, various peaks are visible,
e.g., as indicated by the vertical dashed lines at
$\overh = 0.1591$
and
$\overh = 0.1991$.
These peaks occur in a very complicated manner and
have a strong variation of their widths.
Furthermore, between the peaks also a drastic decrease of
the weight  $\weight{0}$ is observed, e.g.\ for $1/h \approx 0.146$.
Apart from peaks and suppression also plateaus of nearly constant
weight over some interval of $1/h$ are present
for example around the vertical dashed line
at $\overh = 0.375$.

In \prettyref{fig:double_res} Husimi representations
and the corresponding action-space representations in
normal and logarithmic scale
of $\ket{\psi_{\bm 0}}$ are shown for different values of $1/h$.
For the nonresonant case at $1/h = 0.375$,
see  \prettyref{fig:double_res}(a), the state
$\ket{\psi_{\bm 0}}$ localizes
mainly on the quantizing torus $\vecI_{\bm 0}$ (white line).
The Husimi representations for
$1/h = 0.1991$
and
$1/h = 0.1591$
in \prettyref{fig:double_res}(b, c)
show resonant eigenstates with a significant
weight in dynamically distinct regions.
Both cross the resonance junction, however,
the morphology of the states differs, best seen in the
action-space representation, which can be explained
in terms of perturbation theory, which is discussed next.

In order to explain the underlying mechanism of
resonance-assisted tunneling in the case of a double resonance
we use a perturbative approach.
As in Sec.~\prettyref{subsec:scr} we start with the unperturbed system,
but also consider the second resonant term
allowing for two non-vanishing matrix elements
\begin{align}
&\Ibasebra{\bm m} \potentialop \Ibaseket{\bm m + \bm r} =
\Vr, \label{eq:double_coupling_r} \\
& \Ibasebra{\bm m} \potentialop \Ibaseket{\bm m + \bm s} =
\Vs, \label{eq:double_coupling_s}
\end{align}
for quantum numbers differing by  $\pm \bm s$ and  $\pm \bm r$.
The perturbative expansion
developed in Ref.~\cite{Loe1951} and previously similarly applied
to resonance-assisted tunneling in the \twoD{} case
with multiple rank-1 resonances \cite{LoeBaeKetSch2010, Loe2010} reads
\begin{equation}\label{eq:doubleres_pt_state}
\ket{\psi^{\text{pert}}_{\bm m}} =
\Ibaseket{\bm m} +
\sum_{(k, l) \in \mathbb{Z}^2\setminus\{\bm 0\}}
A_{\bm m}^{(k, l)}
\Ibaseket{\bm m + k \bm r + l \bm s}
\end{equation}
for the state localizing predominantly on the quantizing torus
$\vecI_{\bm m}$.
As \prettyref{eq:double_coupling_r} and \prettyref{eq:double_coupling_s}
suggest, there are twice the number of possibilities of subsequent
couplings in each order of perturbation
theory compared to the case of the single resonance.
Thus, in the perturbative expansion of
$\ket{\psi^{\text{pert}}_{\bm m}}$, an unperturbed state
$\Ibaseket{\bm m + k \bm r + l \bm s}$ contributes with a coefficient obtained
by all paths \tunpath{} which connect the initial quantizing torus
$\vecI_{\bm m}$
with the final torus $\vecI_{\bm m + k \bm r + l \bm s}$
in the discrete action grid,
where subsequent actions differ in their quantum numbers
by $\pm \bm r$ or $\pm \bm s$, respectively.
Let $\mathcal{M}_{\bm m}^{k, l}$ be the set of these paths.
In contrast to Sec.~\prettyref{subsec:scr} here every path
$\tunpath \in \mathcal{M}_{\bm m}^{k, l}$ corresponds to a sequence
$\tunpath = \left[\bm t_1, \bm t_2, \ldots, \bm t_{|\tunpath|} \right]$ with
$\bm t_i \in \{ \pm \bm r, \pm \bm s \}$.
Again, paths which return to $\vecI_{\bm m}$ at some point are excluded from
$\mathcal{M}_{\bm m}^{k, l}$.
Taking all paths into account, the coefficients
in \prettyref{eq:doubleres_pt_state} read
\begin{equation}\label{eq:doubleres_pt_coeffs}
A_{\bm m}^{(k, l)} =
\sum_{\tunpath \in \mathcal{M}_{\bm m}^{k, l}}
\lambda_{\bm m}^{\tunpath}
\end{equation}
and every path contributes with
\begin{equation}\label{eq:contrib_path}
\lambda_{\bm m}^{\tunpath}  =
\prod_{i=1}^{\abs{\tunpath}}
\frac{V_{\bm t_i}}{ \hnode \left(\vecIm \right) -
    \hnode \left( \vecI_{\bm m + \sum_{j=1}^{i} \bm t_j }  \right) },
\end{equation}
where we define $V_{-\bm t} = V_{\bm t}$ for
$\bm t \in \{\pm \bm r, \pm \bm s \}$.

Inserting the obtained state $\ket{\psi^{\text{pert}}_{\bm m}} $
of \prettyref{eq:doubleres_pt_state}
into \prettyref{eq:tunneling_rate},
and using the definition of the projection operator \Pabs{},
see \prettyref{eq:def_absorber}, we find
\begin{equation}\label{eq:doubleres_pt_rate}
\weight{m} =
\sum_{\substack{(k, l) \in \mathbb{Z}^2 \\
    \vecI_{\bm m + k \bm r + l \bm s} \in  \psabs} }
    \Big | A_{\bm m}^{(k, l)} \Big |^2.
\end{equation}
In order to apply \prettyref{eq:doubleres_pt_rate} we consider
only the shortest paths \tunpath{}, which reach
\psabs{} as they give rise to the lowest order contributions.
Let $n$ be the length of these paths.
For our choice of the resonance vectors $\bm r$ and $\bm s$,
this restricts the relevant endpoints
$\vecI_{\bm m + k \bm r + l \bm s}$ to those, where $k + l = n$
for positive $k$ and $l$.
Given $(k, l) \in \mathbb{N}^2$ the set
$\mathcal{M}_{\bm m}^{k, l}$ then contains
$\binom{n}{l}$ of these shortest paths.
Here, taking only the shortest paths into account,
i.e.\ the lowest order of perturbation theory, gives
excellent agreement with the weights obtained from numerical diagonalization.
This can be seen in \prettyref{fig:rates_double_res} where the
perturbative prediction of $\weight{0}$,
\prettyref{eq:doubleres_pt_rate}, shown as blue crosses,
is compared with the numerically obtained weights.
The perturbative description covers the overall exponential
decay as well as the peaks and plateaus.
Note that, in general,
considering only the shortest paths may not be sufficient.
One such example is the peak near $\overh = 0.5751$.
As in the case of the single resonance the peaks arise
whenever an intermediate state along one or more perturbative
paths is resonant with $\vecI_{\bm m}$.
Furthermore, quantization jumps occur when the minimal length of
paths, which reach \psabs{}, increases by one.

\begin{figure}
    \includegraphics{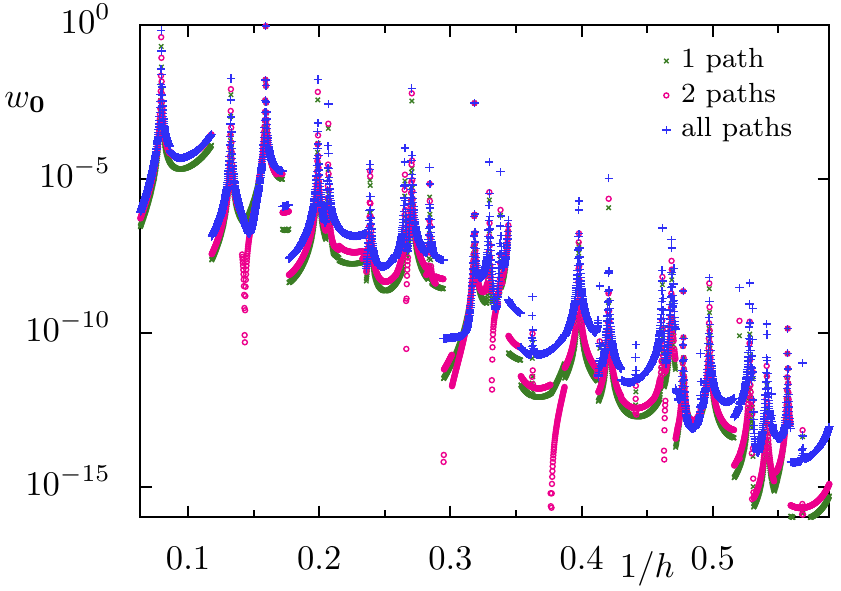}
    \caption{Perturbative prediction using \prettyref{eq:doubleres_pt_rate}
        incorporating an increasing number of paths
        sorted by the absolute value of their contribution
        $\abs{\lambda_{\bm m}^{\tunpath}}$.
        Prediction for one path is shown as crosses,
        for two paths as circles, and for all paths as squares.
       }
    \label{fig:doubleres_pt_paths}
\end{figure}
\begin{figure*}
    \includegraphics{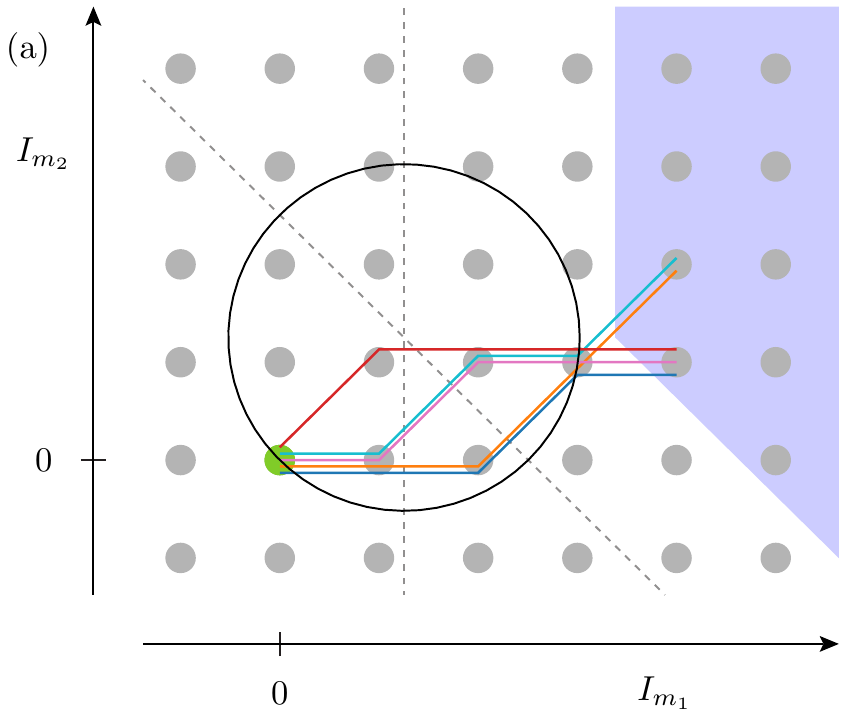}
    \hspace{0.5cm}
    \includegraphics{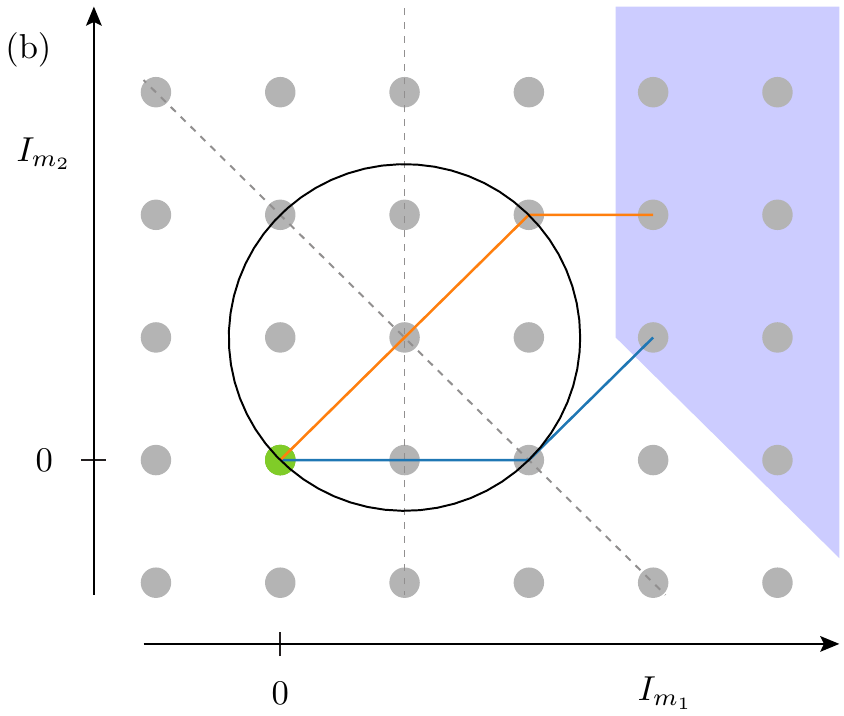}\\
    \vspace{1cm}
    \includegraphics{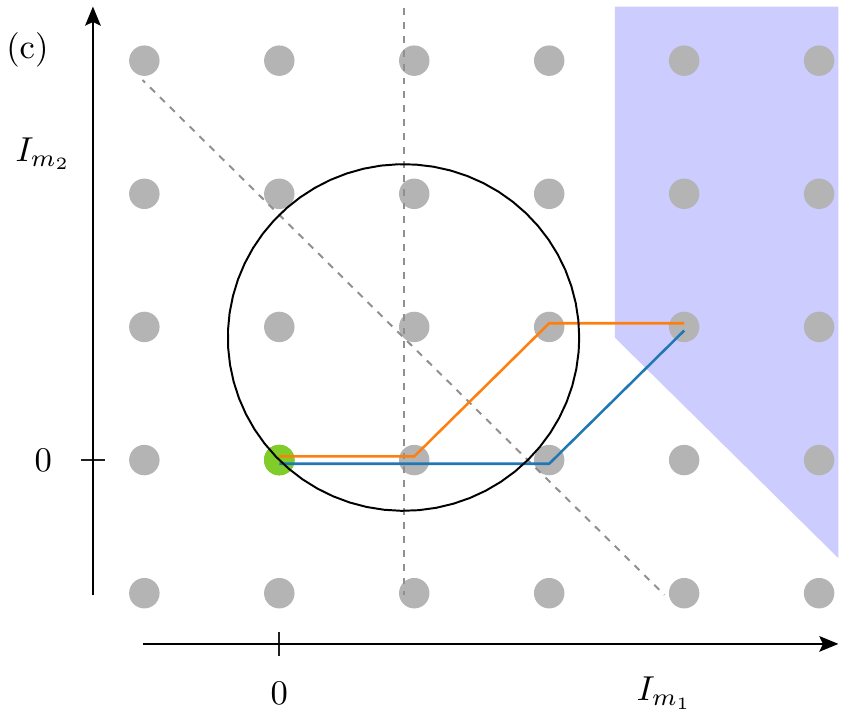}
    \hspace{0.5cm}
    \includegraphics{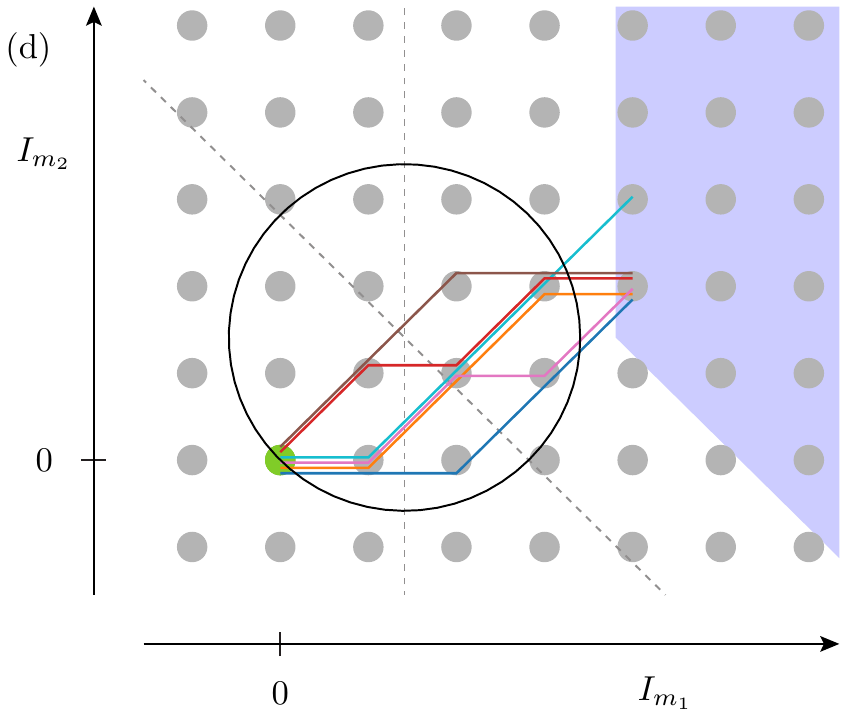}
    \caption{Scheme of perturbation theory for the double
        resonance.
        The action grid \( \vecIm \) is depicted
        as gray points and the
        quantizing torus $\vecI_{\bm 0} = \left( 0, 0 \right)$
        as green point.
        The dashed lines shows the position of the two
        resonance center line,
        see \prettyref{eq:resonance_line}.
        The circle indicates the level set of constant energy
        $\hnode(\vecI_{\bm 0})$.
        The blue shaded area shows the region \psabs{},
        see \prettyref{eq:double_res_abs}.
        In (a) for $\overh = 0.1991$
            five paths with an resonant intermediate state and
        in (b) for $\overh = 0.1591$ two resonant paths with different
                intermediate state are shown.
        In \prettyref{fig:double_res}(b, c)
        both states are shown in a Husimi representation.
        In (c) for $\overh = 0.1465$ two canceling paths are shown
        in (d) for $\overh = 0.2244$ six non-resonant paths
        are shown.
    }
    \label{fig:doubleres_action}
\end{figure*}
In contrast to the single resonance case, there is a larger number of
shortest paths which enter the prediction~\eqref{eq:doubleres_pt_rate}.
As all these paths have the same length, it is a priori not clear which of them
will give the dominant contribution and which paths need to be taken
into account to achieve the accuracy presented in
\prettyref{fig:rates_double_res}.
To this end we sort all contributing paths by their absolute value
$\abs{\lambda_{\bm m}^{\tunpath}}$ in descending order.
In \prettyref{fig:doubleres_pt_paths} the perturbative result
is shown for an increasing number of shortest paths \tunpath{}.
Using only the path with the greatest absolute value for a given
inverse Planck's constant $1/h$ is shown as crosses.
While the positions of the peaks are already resolved,
the plateau-like structures show a mismatch of several orders of magnitude.

In general there is not necessarily one single dominating path,
but several paths may give rise to similar contributions.
This case is illustrated in \prettyref{fig:doubleres_action}(a)
for $\overh = 0.1991$
where on the action grid five
paths containing the resonant intermediate state
$\vecI_{(3, 1)}$
fulfilling
$\hnode(\vecI_{\bm 0}) = \hnode(\vecI_{(3, 1)})$
are shown.
All these paths contribute to $\weight{0}$
within the same order of magnitude.
Thus, considering only the most contributing path resolves the position
of the peaks while adding all resonant paths refines the prediction.
The sixth most contributing path, however, gives rise to a contribution
which is two orders of magnitude smaller since no intermediate state is
located on the circular level set of constant energy $\hnode(\vecI_{\bm 0})$.
Therefore, excluding these nonresonant paths (if other resonant paths
exist) does not affect the result.

In
\prettyref{fig:doubleres_action}(b), two points of the
action grid $\vecI_{(2, 0)}$ and $\vecI_{(2, 2)}$ are close to the level
set of energy $\hnode(\vecI_{\bm 0})$ for ${\overh = 0.1591}$. Thus, both
paths with a resonant intermediate state are essential for an accurate
prediction of the weight.
This also nicely explains the different morphologies
of the Husimi-distributions
and action-space representations,
see \prettyref{fig:double_res}(b, c).
In \prettyref{fig:double_res}(c),
the state $\psiketi$ has a higher density
at the position of both resonant intermediate states
$\vecI_{(2, 0)}$ and $\vecI_{(2, 2)}$.
In contrast, in \prettyref{fig:double_res}(b) only
the intermediate state $\vecI_{(3, 1)}$ shows a higher density.

In contrast to the case where several paths with resonant
intermediate states
constructively interfere also the case of destructive interference
of tunneling paths occurs.
This is best seen in the perturbative prediction
arising from two paths depicted in
\prettyref{fig:doubleres_pt_paths} as circles.
For specific values of \overh{}, e.g.\ around $1/h =0.1465$, a drastic decrease
of $\weight{0}$ can be observed.
This happens if the two paths $\tunpath_1$ and $\tunpath_2$ under
consideration fulfill
$\lambda_{\bm m}^{\tunpath_1} = - \lambda_{\bm m}^{\tunpath_2}$.
\prettyref{fig:doubleres_action}(c) illustrates this scenario,
where two paths cancel each other.
The two involved paths differ in their intermediate
step $\vecI_{(2, 0)}$ and $\vecI_{(2, 1)}$, respectively.
These states lie on different sides of the level set of constant
energy such that the denominator in \prettyref{eq:contrib_path}
differs in its sign but is of equal absolute value.
Taking only the two most contributing paths into account thus gives
a qualitative description of the observed suppression of tunneling.
Such mechanism is also known from \twoD{} maps
if several single resonances are present
\cite{LoeBaeKetSch2010, SchMouUll2011}.
The basic features, i.e.\ the overall exponential decay, the peaks, and
the suppression of tunneling, are already captured if only two paths are
taken into account.
Whereas considering all shortest paths compensates the effect of destructive
interference of paths as depicted in \prettyref{fig:doubleres_pt_paths}.
Furthermore, it emphasizes the necessity to consider all paths
for a quantitatively accurate prediction of the
plateau-like structures pronounced in the regime of large \overh{}.
Schematically, the importance for considering all paths
is depicted in \prettyref{fig:doubleres_action}(d)
showing only the first six most contributing nonresonant paths
for $\overh = 0.2244$.
By considering all paths, the weight of plateau like structures
can be quantitatively predicted.

Note that for arbitrary resonance vectors,
higher orders of perturbation theory
may be necessary for instance if there are
paths with a resonant intermediate state
which need an additional step to reach \psabs{}.
In other words, the shortest paths do not necessarily lead to the largest
contribution.
If necessary, these longer paths can be computed easily to refine
the prediction.
In the considered system, including only the shortest path
in the perturbative expression \eqref{eq:doubleres_pt_rate}
allows for a successful prediction of the
numerically obtained weight $\weight{0}$, see \prettyref{eq:tunneling_rate},
over several orders of magnitude.

\subsection{\fourfour{} matrix model}\label{subsec:matrixmodel}

\begin{figure}[t]
    \includegraphics{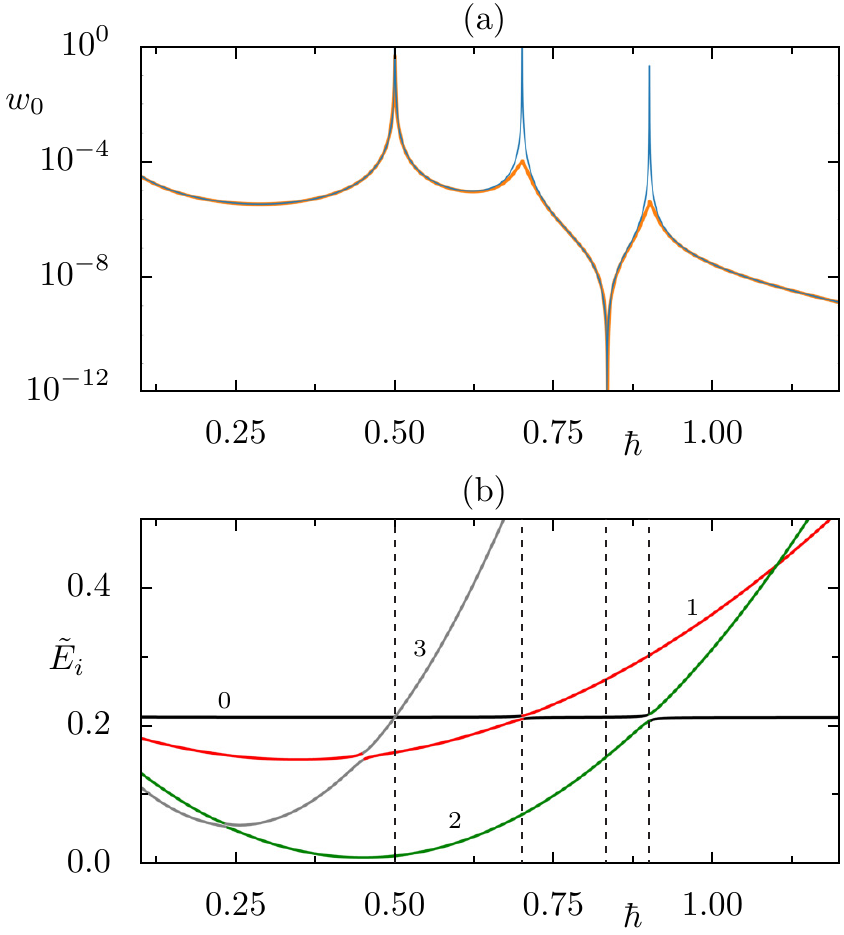}
    \caption{Matrix model:
        (a) Weight $w_0$ as a function of $\hbar$
        obtained via \prettyref{eq:block_gamma} shown as orange line
        and $w_{0, \text{pert}}$ calculated using \prettyref{eq:block_pert}
        as blue line.
        (b) Energies $\tilde{E}_i$ (solid lines) of the
        states  \( \ket{\psi_i} \)
        as a function of $\hbar$ in black, red, green and
        gray for $i=0,1,2,3$.
        The vertical dashed lines indicate the values of $\hbar$ for
        the cases of enhancement and suppression.
    }
    \label{fig:block_model}
\end{figure}

The underlying mechanism of the enhancement and suppression of
the weight $\weight{m}$ in the situation of a double resonance
can be explained within a minimal \fourfour{} matrix model.
To this end, we consider four states $\Ibaseket{i}$ resembling
 the unperturbed action states
\( \Ibaseket{\bm m}\)
corresponding to the quantizing actions
$\vecI_{0} = \vecI_{(0, 0)} = (0, 0) $,
$\vecI_{1} = \vecI_{(1, 0)}  = (\hbar, 0) $,
$\vecI_{2} = \vecI_{(1, 1)} =(\hbar, \hbar) $ and
$\vecI_{3} = \vecI_{(2, 1)}  = (2 \hbar, \hbar)$.
The unperturbed energies of these states are given by
$ E_{i} = \hnode (\bm I_i) $, see \prettyref{eq:scr_cl_ham_h0}.
We allow for couplings between these states according to
\prettyref{eq:double_coupling_r} and \prettyref{eq:double_coupling_s}.
The matrix representation then reads
\begin{equation}\label{eq:block_matrix}
\Hblock =
\begin{pmatrix}
E_{0} & \Vr & \Vs & 0 \\
\Vr & E_{1} & 0 & \Vs \\
\Vs & 0 & E_{2} & \Vr \\
0 & \Vs & \Vr & E_{3}
\end{pmatrix}.
\end{equation}
Diagonalization of $\Hblock$ for a fixed value of $\hbar$ yields the
eigenstates
$\ket{\psi_i}$ in the basis of action eigenstates.
We choose for $\hnode$ the parameters $M_{\text{res}} = 1$ and
$\vecIs = (0.35, 0.55)$ and for the resonances
$\bm r = (1, 0)$ and $\bm s = (1,1) $
with corresponding coupling strengths $\Vr{} = 0.0025$ and $\Vs{} = 0.005$,
respectively.
As the perturbations are small, the eigenstates will predominantly
resemble one of
the action states and are labeled accordingly.
Thus, resonance--assisted tunneling between $\vecI_0$ and $\vecI_3$ can
be quantified by the overlap
\begin{equation}\label{eq:block_gamma}
w_0 = \abs{ \braket{\psi_0|\vecI_3} }^{2}
\end{equation}
between the state $\ket{\psi_0}$ associated with $\vecI_{0}$
and the state $\Ibaseket{3}$.
The states $\Ibaseket{1}$ and $\Ibaseket{2}$ deal as intermediate states.
In \prettyref{fig:block_model}(a)
the weight $w_0$ is depicted as a function of
$\hbar$ as (orange) line.
It resembles the basic features of resonance-assisted
tunneling observed in the full system with a double resonance.
In particular, the \fourfour{} matrix models shows three peaks of enhancement
as well as suppression in between the second and third peaks.

In the following, we treat the \fourfour{} matrix model also perturbatively.
In order to compute the
weight $w_0$, \prettyref{eq:block_gamma}, there are only two
paths to be considered.
They connect $\vecI_{0}$ with $\vecI_{3}$ and differ by
their intermediate step.
Explicitly they are given by \( \tunpath_1 = [\bm r, \bm s] \) and
\( \tunpath_2 = [\bm s, \bm r] \)
and contribute with
\begin{align}
&\lambda_1 = \frac{\Vr}{E_0 - E_1}
\frac{\Vs}{E_0 - E_3}, \label{eq:block_path_one}\\
&\lambda_2 = \frac{\Vs}{E_0 - E_2}
\frac{\Vr}{E_0 - E_3}. \label{eq:block_path_two}
\end{align}
The perturbatively computed weight is given by
\begin{equation}\label{eq:block_pert}
w_{0, \text{pert}} = \abs{\lambda_1 + \lambda_2}^2
\end{equation}
and is shown as the (blue) line in \prettyref{fig:block_model}(a).
They are in perfect agreement with the numerically obtained weights
away from the peaks and
as long as the coupling terms are sufficiently small.
The position of the peaks are given by the denominators
of \prettyref{eq:block_path_one} and \prettyref{eq:block_path_two},
whenever there is a degeneracy of the unperturbed state $\Ibaseket{0}$
with either one of the remaining three states.
This is exactly the case for
$ \hbar = 2 \Irsone $, $\hbar = \Irsone + \Irstwo$ and
$ \hbar = \frac{4\Irsone + \Irstwo}{5}$.
Instead, destructive interference
occurs if $\lambda_1 = -\lambda_2$ which is equivalent to
 $(E_0 - E_1) = -(E_0 - E_2)$ and holds for
$\hbar = \frac{4\Irsone + 2\Irstwo}{3}$.
The values of $\hbar$ for enhancement and suppression
are marked in \prettyref{fig:block_model} as black dashed lines.
For the case of suppression, the origin of the opposing signs
can be read off from the denominators
of \prettyref{eq:block_path_one} and \prettyref{eq:block_path_two},
indicating that states $\vecI_{1}$ and $\vecI_{2}$ lie on different
sides of the level set of constant energies  and
have equal energy difference from $E_0$ but with a different sign.

Additionally, in \prettyref{fig:block_model}(b) the energy levels
$\tilde{E}$ of the eigenstates
$\ket{\psi_i}$ are shown.
The black line corresponds to $\tilde{E_0}$ and shows avoided
crossings with either the energies of the intermediate states
(green and red)
or the energy of the final state (gray).
The positions of these avoided crossings match with
the positions of the peaks of enhancement of the weight $w_0$.
On the other hand, in the case of suppression, the energy difference
between $\tilde{E_0}$ and the energy of the two intermediate states is equal
in size, but of opposite sign.

\section{Summary and outlook}\label{sec:summary}

In this paper, we studied resonance--assisted tunneling
by exploiting the universal description of the classical dynamics
in the vicinity of a nonlinear resonance in terms of normal--form Hamiltonians.
In particular we concentrated on single as well as
double resonances in \fourD{} normal--form Hamiltonians,
where we visualized classical phase space in a suitable
hyperplane.
Numerical diagonalization of the quantized normal--form Hamiltonians
yielded its
eigenstates, whose Husimi representation allowed for the comparison with
classical phase-space structures.
There, we observed enhanced probability on classical tori
located on opposite sides of the resonance channels.
In particular, for specific
values of the effective Planck's constant, this effect becomes significantly
enhanced due to resonance--assisted tunneling.
We introduced the weight of an eigenstate localizing on one side
of the resonance channel in a disjoint phase-space region as a quantitative
measure of tunneling,
which gives qualitatively the same behavior as, e.g., phase splittings or
tunneling rates studied in a \twoD{} system.
That is, we found an overall exponential decay as well as prominent peaks,
where tunneling is enhanced over several orders of magnitude,
showing a complicated peak structure
already for just one double resonance.
Furthermore in this situation suppression of tunneling is possible.
This also occurs in \twoD{} systems, but only when multiple
single resonances are involved, which happens
in the deep semiclassical regime in which also
small resonance chains become important.
This is different in \fourD{} systems, as typically double resonances
dominate resonance--assisted tunneling
even in the quantum regime of large $\hbar$.

In order to predict the weight and understand the mechanism for the peaks
and cases of suppression, the use of the normal--form Hamiltonians
allowed us to perform the perturbative calculation in both the single and the
double resonance case.
Tunneling across the single resonance
turns out to be effectively \twoD{} due to a second constant of motion.
In contrast, the situation of the double resonance is more involved.
This was visualized by introducing paths on
the discrete action grid.
There, for the single resonance, just one path
needs to be considered
while for the double resonance multiple paths must be included to obtain an
accurate description.
In particular, this allows for destructive interference of different paths
leading to the suppression of tunneling for specific values of $\hbar$.
In contrast, the mechanism causing the peaks is the same for both
cases and corresponds with the one known from \twoD{} systems.
Crucial for enhancement in either case is the energetic degeneracy of action
states with respect to the unperturbed Hamiltonian.
Note that for different sets of resonance vectors
even higher orders of perturbation theory may yield
the dominant contribution in comparison with lower orders,
i.e.\ shorter paths.
For the concrete systems studied in this paper,
this was not the case and it was
sufficient to take only the lowest order in terms of the shortest paths into
account.
Furthermore, we presented a minimal \fourfour{} matrix model, which captures
the observed features of resonance--assisted tunneling in \fourD{}
and allowed for a simplified explanation of the observed phenomena.

The perturbative description
provides a first step toward a detailed understanding
of resonance--assisted tunneling
for general higher--dimensional systems,
e.g.\ for experimentally feasible three-dimensional optical microcavities
or microwave resonators.
For a universal description of a generic system, e.g.\ a \fourD{}
quantum map, several presently open problems have to be solved:
For example, in phase space the relevant resonances for
a given regime of $\hbar$ need to be identified.
For these resonances the parameters for the construction
of the normal--form Hamiltonian have to be extracted.
Furthermore, for these local approximations of the resonances,
a canonical transformation to the phase space of the map
needs to be constructed.
The subsequent quantization would then allow for a quantitative
prediction of the tunneling rates and resonance-assisted tunneling peaks.
Beyond that, the development of a semiclassical description only based on the
classical properties of the nonlinear resonance would be desirable.
Another phenomenon, occurring in at least \fourD{} symplectic maps or
\sixD{} Hamiltonians, is the famous Arnold diffusion.
It provides a classical transport mechanism
connecting different regions in phase space.
Its interplay with tunneling is not clear at present.

\begin{acknowledgments}
  We are grateful for discussions with
  Martin Richter and Normann Mertig.
  Furthermore, we acknowledge support by the Deutsche Forschungsgemeinschaft
  under grants BA~1973/4--1 and KE~537/6--1.

  All \threeDD{} visualizations were created using
  \textsc{Mayavi}~\cite{RamVar2011}.

\end{acknowledgments}

\end{document}